\documentclass[preprint]{elsarticle}

\usepackage{amssymb}
\usepackage{amsmath}
\usepackage{array, longtable, tabularx}% added long table
\usepackage{color}

\usepackage{lineno}
\journal{Entertainment Computing}

\begin{document}

\begin{frontmatter}

\title{Emotion Dynamics in Social Deception Games: Analysis of Professional and Nonprofessional Players through Electrodermal Activity in Werewolf Games}

\author[1]{Sho Mitarai\corref{cor1}} 
\author[1]{Chang Liu}
\author[1]{Goshiro Yamamoto}
\author[2]{Nagisa Munekata}

\cortext[cor1]{Corresponding author}

\affiliation[1]{organization={Kyoto University},%Department and Organization
            addressline={53 Shogoin Kawaracho, Sakyo Ward}, 
            city={Kyoto},
            postcode={606-8507}, 
            state={Kyoto},
            country={Japan}}
\affiliation[2]{organization={Kyoto Sangyo University},%Department and Organization
            addressline={Kamigamo motoyama, Kita Ward}, 
            city={Kyoto},
            postcode={603-8555}, 
            state={Kyoto},
            country={Japan}}

%% Abstract
\begin{abstract}
The development of AI systems capable of emotionally resonant communication remains a significant challenge. This study examines how humans influence emotions in social deception games by comparing professional and non-professional players. We measured electrodermal activity during gameplay to capture physiological emotional responses and analyzed communication patterns during periods of high emotional arousal. Our results revealed distinct communication strategies: professional players maintained persuasion-based approaches under high arousal, while nonprofessional players shifted toward information-focused communication. Statistical analysis confirmed significant differences in expression patterns between expertise levels. Professional players exhibited more stable emotional states during gameplay, indicating better emotional regulation. These findings inform the design of AI systems that can adapt their communication strategies based on recipient characteristics, advancing the development of emotionally intelligent artificial agents.

\end{abstract}

%% Keywords
\begin{keyword}
emotion dynamics \sep werewolf game \sep social deception game \sep electrodermal activity 


\end{keyword}

\end{frontmatter}

\section{Introduction}
As artificial intelligence systems are increasingly integrated into our daily lives, the ability to engage in emotionally resonant communication has become a crucial challenge~\cite{Werewolf-XL}. Although AI has demonstrated remarkable success in various domains, developing systems that can emotionally engage humans through communication remains a significant challenge, particularly in complex social situations involving trust, deception, and emotional manipulation.

Among various contexts where emotional and strategic communication plays a crucial role, social deception (deduction) games, particularly "Werewolf game," provide an ideal environment for studying such interactions. In Werewolf, players must carefully select their communication strategies to achieve their goals while dealing with incomplete information and potential deception. While projects like AI Wolf have advanced the development of game-playing agents, these systems still struggle to engage in communication that effectively influences human emotions within complex emotion dynamics \cite{EmotionDynamics}. This limitation stems in part from the challenge of understanding genuine human intentions, which are often concealed beneath surface-level expressions and statements.

In human communication, people employ various strategies to convey their intentions and influence others, ranging from logical persuasion to straightforward information sharing. The effectiveness of these strategies depends on both the speaker's skill level and the recipient's ability to comprehend and process the information. This fundamental aspect of communication leads us to hypothesize that in social deception games, players of different skill levels might employ and respond to different communication strategies. In addition, the emotional responses to game events and interactions may vary depending on players' experience and strategic understanding. However, examining these differences presents a methodological challenge, as social deception games inherently involve concealing true intentions and emotional states.

To address this challenge in understanding genuine emotional responses, we employ electrodermal activity (EDA) measurements, which can detect physiological responses that are difficult to consciously control. This approach is particularly valuable in social deception games, where players, especially those with expertise, often masterfully control their outward expressions and verbal behaviors to mislead others while maintaining hidden strategies. Our study specifically investigates two key aspects: (1) the physiological differences between professional and nonprofessional players, revealing how expertise influences emotional regulation during deceptive interactions, and (2) the relationship between specific verbal interactions and emotional responses, uncovering what types of communication affect players' emotional states regardless of their outward reactions.
The contributions of this study are as follows.
\begin{enumerate}
    \item We present the first empirical evidence of how expertise influences communication strategies and emotional responses in social deception games through physiological measurements. This analysis reveals differences in how professional and nonprofessional players handle high arousal situations.
    \item We identify specific patterns in communication strategies that correspond to different levels of expertise, showing how players' logical thinking abilities and emotional resilience shape their approach to persuasion and information sharing.
\end{enumerate}
These insights are particularly valuable for entertainment computing, as they provide a deeper understanding of player behavior in social deception games. Our analysis of expertise-dependent communication strategies helps illuminate the complex social dynamics that make such games engaging. Furthermore, our findings offer preliminary insights into how humans adapt their communication approaches based on emotional states, which extends beyond gaming contexts to various social interaction scenarios where strategic communication plays a crucial role.

\section{Related Work}
\subsection{Game AI and Incomplete Information Games}
Recent advances in artificial intelligence (AI) have demonstrated superhuman performance in both complete information games and some types of incomplete information game. While complete information games like chess and Go have been effectively conquered by AI through deep learning and search techniques~\cite{AlphaGo}, incomplete information games like poker have required additional innovations to handle hidden information and opponent modeling~\cite{SuperhumanPokerAI}. However, social deception games like the werewolf game present unique challenges that go beyond traditional AI approches.

The AIWolf Project~\cite{AIWolf1, AIWolfProject1, AIWolfProject2} has pioneered research on AI agents for the werewolf game, highlighting the complexity of modeling human-like behavior in social deception games. Prior research has shown that the Werewolf game requires players to estimate other players' beliefs and intentions under incomplete information, necessitating a formal model that can handle probabilistic mental states.~\cite{ExtendedBDIModel}. Unlike poker, where probability calculations can guide optimal play, the werewolf game involves complex social dynamics that current AI systems struggle to master~\cite{HumanLikeWerewolfAgents}.

\subsection{Analysis of Player Behavior in Werewolf Games}
Previous reseach on player behavior in werewolf games has focused primarily on observable aspects of gameplay. Linguistic analysis studies have examined patterns in persuasive arguments, deceptive statements, and role-specific languagge use~\cite{DevelopWerewolfAgentswithQLearning, WerewolfModelBasedOnPlayLogAnalysis}. These studies have identified certain verbal patterns associated with different player roles and game outcomes but are limited by players' ability to intentionally control their speech.

Non-verbal behabior analysis has investigated facial expressions~\cite{AnalysisOfNonVerbalInformation}, body language~\cite{AnalysisOfNonVerbalInformation, AnalysisOfNonVerbalInformationForAgentDesign, WerewolfAnalysisWithGazeMotion} during gameplay. While these studies have found some correlations between observable behaviors and player roles or intentions, they face a limitation: experienced players can consciously manipulate these signals to deceive others, making such surface-level analysis potentially unreliable.

\subsection{Phsiological Measurements in Gaming and Communication Research}

Research on physiological responses during social interactions has shown that EDA can be a reliable indicator of arousal during deceptive behavior~\cite{DeceptiveBehaviorWithEDA} and decision-making~\cite{DecisionMakingWithEDA}. 

Studies in interpersonal communication have demonstrated how physiological responses can reveal emotional states that are not apparent through behavioral observation alone~\cite{InterpersonalAutonomicPhysiology}.
However, research on psysiological responses in social deception games remains limited, particularly in games like werewolf that require analysis of multiple players' psysiological responses. Previous studies using physiological measurements in games have focused mainly on single-player experiences~\cite{EDAAnalysisInFPSGame, AffectiveGaming} or competitive games without social deception elements.

Our research addresses this gap by examining how physiological responses differ between professional and nonprofessional werewolf game players, and how these responses correlate with specific verbal interactions. Specifically, we investigate how the players' emotion dynamics interact with each other by simultaneously measuring the EDA of all players in a five-player werewolf game. Furthermore, we difine professional players—a group that has not been focused on in previous werewolf game research—and analyze their behavioral patterns, emotional states, and verbal communication in social deception interactions by comparing them with nonprofessional players.

\section{Experiments}
We conducted experiments measuring players' electrodermal activity (EDA) during werewolf games. Each experiment involved a 5-player werewolf game and recorded video, audio, and EDA data. The experiments with professional and nonprofessional players were conducted at different times and locations. We note any differences in experimental conditions between the two participant groups where relevant.

\subsection{Participants}

\subsubsection{Professional}
Seven actors (five male, two female) from
% "Jinrou The Live Playing Theater"
[anonymized for blind review], a theater company that produces drama performances with werewolf game, participated in this study. These participants professionally perform and play werewolf games as part of their theatrical productions, with approximately 150-400 performances each. Additionally, they regularly play werewolf games as part of their theatrical training, and report experience with over 1,000 game sessions. Based on these criteria, the participants were considered to be highly skilled players and were classified as professionals in this study. While their professional work primarily involves 13-player werewolf games, the 5-player version used in this experiment differs in role types, game progression, and standard strategies. Therefore, participants completed approximately 30 practice sessions of 5-player werewolf games beforehand to ensure thorough understanding of the game progression and standard strategies before conducting the experiment. This study was approved by the Ethics Committee for Research Involving Human Subjects at [anonymized for blind review]. 

\subsubsection{Nonprofessional}
Six participants (all male) were selected based on their prior gameplay experience of "Werewolf." The participants regularly played the five-player version of the werewolf game and were not novice, possessing sufficient understanding of the rules, similar levels of experience, and familiarity with the game. This study was approved by the Research Ethics Committee for Human Subjects at [anonymized for blind review]. %the University of Tsukuba.

\subsection{Experimental Setup}
\subsubsection{Game Settings}
The werewolf game used in this experiment is a 5-player version. The five players are divided into the villager team (3 players) and the werewolf team (2 players). The roles consist of two villagers, one seer, one madman, and one werewolf. The details of each role are shown in Table \ref{tab:RoleDescription}.  Roles were assigned to players in a balanced manner throughout the experiment to avoid bias. Since there were more than five participants, players alternated to ensure equal participation.

\begin{table}[]
    \caption{Roles, abilities, and number of players (N) assigned to each role in five-player werewolf game.}
    \centering
    \begin{tabularx}{\textwidth}{l|X|c}
    \hline
        Role & Ability & N \\ \hline
        Villager & A member of the village team without special abilities. Must identify and eliminate werewolves through discussion and voting. & 2\\ 
        Seer & A village team member who can investigate one player per night phase to determine if they are a werewolf or not. & 1 \\
        Madman & A member of the werewolf team member who appears as a villager to the Seer's investigation. Has no special abilities but wins with the werewolf team. & 1\\ 
        Werewolf & A werewolf team member who can eliminate one player per night phase (except the first night phase). & 1\\
    \hline
    \end{tabularx}
    \label{tab:RoleDescription}
\end{table}

\begin{figure}[tb]
      \begin{center}
      \includegraphics[width=\textwidth]{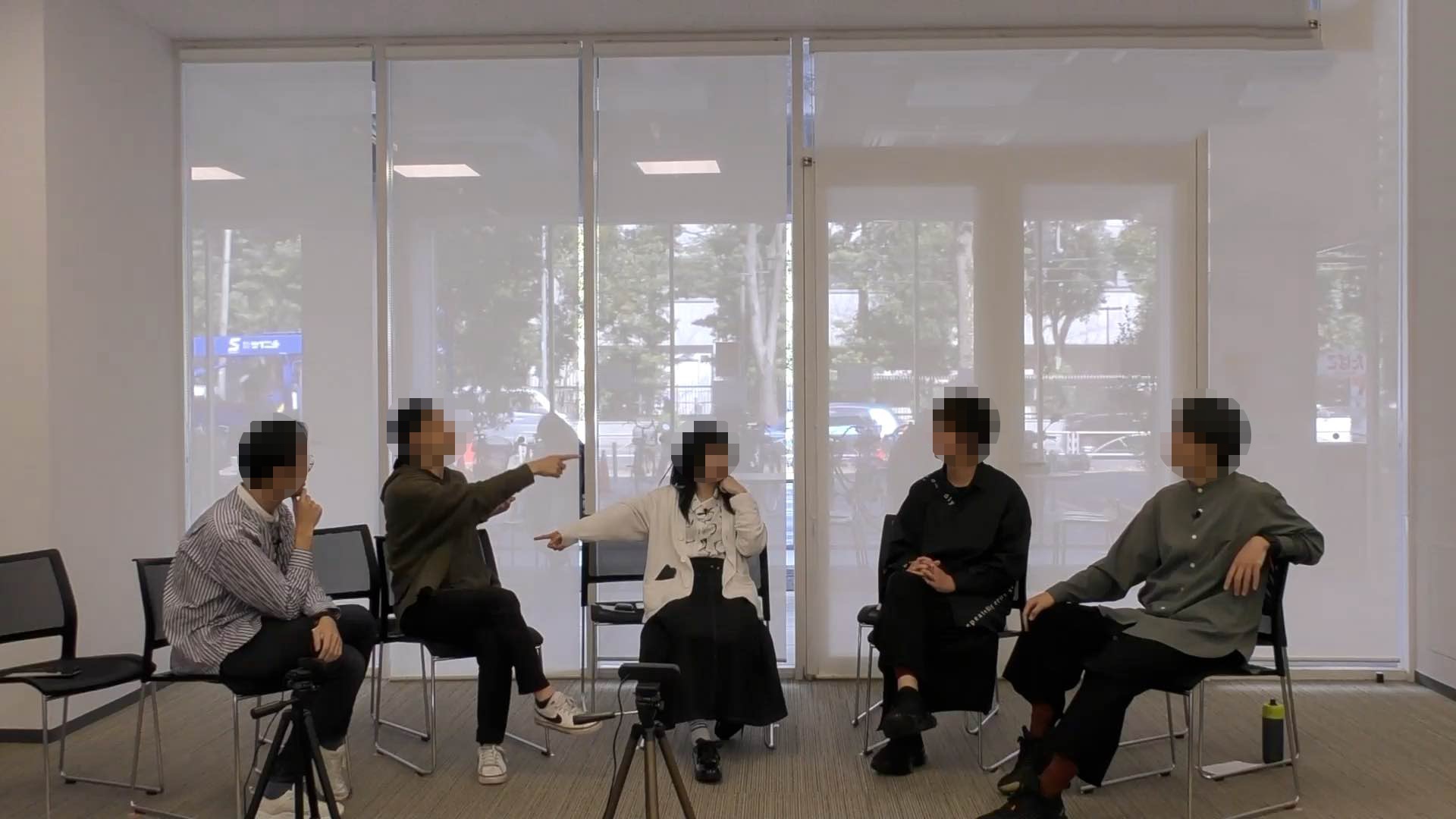}
      \caption{Experimental setup for the werewolf game. Five players are seated in a semi-circle to facilitate face-to-face communication. Each player wears an EDA sensor on their non-dominant hand to measure skin conductance level during gameplay.}
      % \ecaption{Experimental Environment.}
      \label{fig:experimental-environment-tlpt}
      \end{center}
\end{figure}

Players participated in the werewolf game while seated in a semicircle as shown in Figure \ref{fig:experimental-environment-tlpt}. The game proceeds as follows: 
\begin{enumerate}
  \item Game Master distributes role cards to players
  \item Night Phase: The Seer views one player's role
  \item Discussion Phase (Nonprofessional players: 4 min.; Professional players: 5 min.)
  \item Voting Phase
  \item Results announcement. Game continues if the werewolf is not executed.
  \item Night Phase: The Seer views one player's role, and the werewolf eliminates one player
  \item Discussion Phase (Nonprofessional players: 4 min.; Professional players: 5 min.)
  \item Voting Phase
  \item Results announcement
\end{enumerate}
The Game Master (GM) distributes roles, and all players confirm their assigned roles. During the Night Phase, the werewolf eliminates a player, and the seer verifies another player's role (werewolf or not). In the Discussion Phase, all players deliberate, followed by the Voting Phase where they select who to execute. During the Voting Phase, players vote for those they suspect of being the werewolf, and the player receiving the most votes is executed. In a tie for the most votes, the outcome is determined through a runoff vote. During the runoff voting, each tied player has one minute to present their defense, while others listen before casting their votes. If the executed player is the werewolf, the villager team wins; if not, the game returns to the Nigh Phase (step 2). In this 5-player version, if the werewolf survives two Voting Phases, the werewolf team wins.

\subsubsection{Apparatus}
A BIOPAC Systems BN-PPGED (for professionals) and our developed Q-sensor-like~\cite{QSensor} (for nonprofessional) were used for EDA measurement. A pair of electrodes were placed to each participant's nondominalt palm (left hand for all participants). Since the electrodes were wirelessly connected to the measurement device, they did not interfere with participants' behaviors, such as gesturing. 

\subsection{Data Analysis}
\subsubsection{EDA Analysis}
\begin{figure}[t]
      \begin{center}
      \includegraphics[width=.8\textwidth]{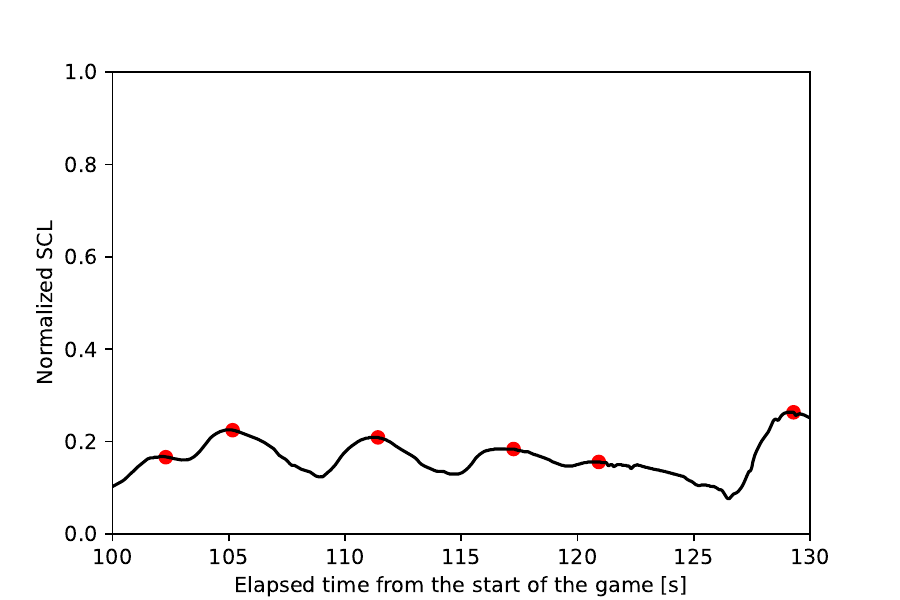}
      \caption{Example of SCR frequency. The graph shows normalized SCL over a 30-second period (100-130 seconds from game start), where red dots indicate identified SCR components. In this example, 6 SCR components were identified over 30 seconds, yielding 12 components per minutes.}
      \label{fig:peak_example}
      \end{center}
\end{figure}

The EDA data were analyzed from two perspectives. The first examines the trend of the overall skin conductance level (SCL). By analyzing the magnitude of the SCL obtained from the experiment, we can understand the general emotional tendencies. The second focuses on the frequency of the skin conductance response (SCR) components. By calculating the peak values from the SCL data obtained, we can determine how frequently arousal responses occurred, providing information about localized excitement. As an example, Figure \ref{fig:peak_example} shows the components of SCL and SCR for a 30-second segment. The red dots in the figure indicate SCR components. For comparison purposes, frequencies were normalized by the duration of each segment to obtain the frequency per minute.

Regarding SCL analysis, due to physiological characteristics, changes in skin conductance vary between individuals, necessitating analysis of interpersonal variations for each participant. In addition, different devices were used to measure the professional and nonprofessional participants in this experiment. To address these individual differences and equipment variations, the sensor values were normalized to a range of 0.0 to 1.0. Although differences in EDA measurements are possible due to equipment performance variations, both devices measure skin conductance activity using the transconductance method, which measures apparent resistance by passing a small current through two electrodes attached to the palm. Therefore, normalization is considered sufficient to resolve these differences. Normalization was performed for each trial, with the normalized SCL value of 0.0 representing the minimum and 1.0 representing the maximum value for that trial.

The analysis focuses on the Discussion and Voting phases of the first day. This focus was selected because: 1) the Discussion and Voting phases of the first day occur in every game (if the game ends on the first day, there are no second-day discussions and voting), allowing for analysis of more trials; and 2) most games were decided on the first day, suggesting that crucial elements of the werewolf game are contained in the first day's discussion. During the Discussion phase in this experiment, we observed consistent patterns in the discussion content throughout the trials, leading us to divide the Discussion phase into four distinct events. Table \ref{table:event} shows an overview of these events.

For statistical analysis, we performed a two-way ANOVA for mixed design with the groups (professional - nonprofessional) and the events (events of the Discussion phase and the Voting phase) on normalized SCL values and SCR frequencies. The significance level was set at $\alpha = 0.05$. When significant main effect or interactions were found, post-hoc comparisons were perfomed using Holm's method with appropriate adjustment for multiple comparisons.

\begin{table*}[t]
\caption{Discussions During the Meeting Phase}
\label{table:event}

\begin{tabularx}{\textwidth}{lXl}
\hline
 &  & Time \\ \hline 
Early & A period of mutual probing. This phase frequently features role revelations where the seer comes forward to announce their divination results. & 14 s - 17 s \\
Deduction & Players make assumptions and deduce roles based on the divination results. During this phase, additional players sometimes came forward claiming to be the seer. & 1 m, 34 s - 1 m, 57 s\\
Doubt & Players express suspicions about who belongs to the werewolf team. The discussion centers on reasons for suspicion and defenses from suspected players. & 1 m, 36 s - 2 m \\
Summary & Players organize all previously discussed information and deliberate on their voting decisions. & 36 s - 45 s \\
\hline
\end{tabularx}
\end{table*}

\subsubsection{Analysis of Utterance Data}
We analyzed the expressions of the players during the werewolf game by categorizing them into five main categories according to their discussion patterns. These categories were derived from preliminary observations of game interactions and reflect different aspects of player communication in social deception games. Table \ref{table:category} shows our categorization framework with main categories and their subcategories.

We analyzed the expressions under two conditions: during maximum SCL peaks and during randomly selected first-day discussions. For the maximum SCL condition, we extracted the utterances made at each players' highest SCL point. For the random condition, we randomly sampled utterances from the first-day discussions of each game to provide a baseline for comparison.

A researcher categorized the utterances according to the framework shown in Table \ref{table:category}. To ensure consistency in categorization, the researcher first conducted a preliminary analysis on a subset of the data to establish clear categorization criteria (see Appendix A). After establishing the categorization approach, they proceeded with the full analysis. The final dataset included 120 utterances (60: max SCL, 60: random) from professional players and 100 utterances (50: max SCL, 50: random) from nonprofessional players.

\begin{table}[tb]
\caption{Classification Categories of Utterances}
\begin{tabular}{l|l}
\hline
Main category        & Subcategory                                       \\ \hline
Persuation/Assertion & Share deduction process                           \\
                     & Agree/Support                                     \\
                     & Disagree/Reject                                   \\
                     & Explain their action/utterance intention          \\
                     & Summarize and persuation                          \\
                     
                     & Point out inconsistencies                         \\ \hline
Information Management          & Provide information                               \\
                     & Role declaration/revocation                       \\
                     & Report seer's result                              \\ \hline
Emotional Statement              & Trust statements                                  \\
                     & Suspicion statements                              \\
                     & Question/Confirmation                             \\
                     & Confirm intent of statement and reason for action \\
                     & Confirm someone's role                            \\ 
                     & Explain a condition of themselves                 \\
                     & Promote agreement                                 \\ \hline
Direction            & Declare where to vote                             \\
                     & Direction of Discussion                           \\
                     & Promote to dicide where to vote                   \\
                     & Game progression                                  \\ \hline
Unclassified/Others  & Unclassified/Others                               \\ \hline
\end{tabular}
\label{table:category}
\end{table}

We conducted Mann-Whitney U tests to examine differences in the distribution of expression categories. The tests were performed for three comparisons: (1) between maximum SCL and random conditions for professional players, (2) between max SCL and random conditions for nonprofessional players, and (3) between professional and nonprofessional players during max SCL periods. The significance level was set at $\alpha = 0.05$. 

\section{Results}
We conducted 16 trials with professional players and 15 trials with nonprofessional players, each trial continuing until a team's victory was determined. To maintain natural postures during gameplay behavior, we did not implement strict movement restrictions. Consequently, participants could move their hands and arms freely, resulting in data loss in two (professional) and five (nonprofessional) trials due to actions such as pressing firmly on the electrode areas. In addition, two trials with professional players were conducted without electrodermal activity (EDA) measurements. Therefore, this study analyzes 12 out of 16 games for professional players and 10 out of 15 games for nonprofessional players.

\subsection{Overall EDA Patterns}
\begin{figure}[tb]
      \begin{center}
      \includegraphics[width=\textwidth]{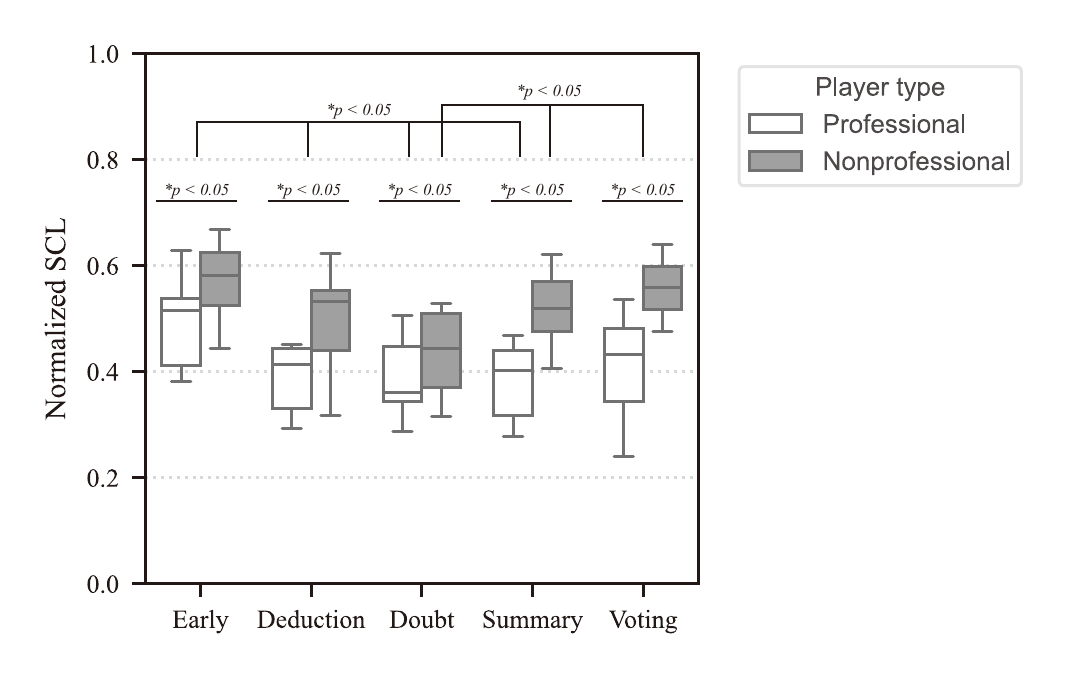}
      \caption{Comparison of normalized SCL values between professional and nonprofessional players across different game events. Statistical significance was determined using two-way mixed ANOVA followed by post-hoc analysis with Holm's method ($*p < 0.05$). Horizontal bars at the top indicate significant differences between events, while asterisks above event pairs show significant differences between player types.}
      \label{fig:scl_trend}
      \end{center}
\end{figure}

Figure \ref{fig:scl_trend} shows the general trends of normalized SCL obtained in this experiment. A two-way mixed ANOVA (Factor 1: professional / nonprofessional [between participants], Factor 2: events [within-participants]) was conducted to examine group relationships, revealing main effects for both factors (Factor 1: [F(1,11)=6.50, $*p<0.05$], Factor 2: [F(4, 11) = 11.09, $**p<0.01$]). Post-hoc analysis using Holm's method for multiple comparisons of factor 2 showed that normalized SCL during the early discussion phase was significantly higher than in all other events except voting ($*p<0.05$). Furthermore, the normalized SCL during the voting phase was significantly higher than during both the suspicion and summary discussion phases ($*p<0.05$). These results indicate that nonprofessional players exhibited higher overall arousal during the discussion phase. Furthermore, arousal levels varied across different phases, with higher levels observed during the early discussion and voting phases.

\begin{figure}[tb]
      \begin{center}
      \includegraphics[width=\textwidth]{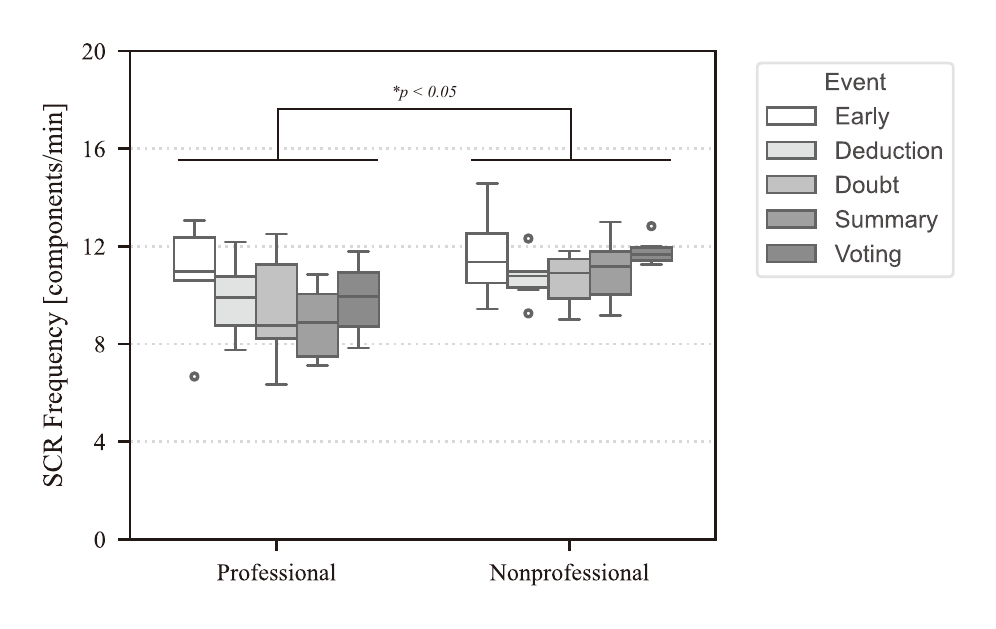}
      \caption{Comparison of SCR frequency (components/min) between professional and nonprofessional players across different game events. Box plots show the distribution of SCR frequencies during Early, Deduction, Doubt, Summary, and Voting phases. Statistical analysis using two-way mixed ANOVA revealed significant differences between player types ($*p < 0.05$), with nonprofessional players showing consistently higher SCR frequencies across all events. Circles represent outliers.}
      \label{fig:scr_peak}
      \end{center}
\end{figure}

Figure \ref{fig:scr_peak} shows the frequency of SCR components. A two-way mixed ANOVA (Factor 1: professional / nonprofessional [between participants], Factor 2: events [within participants]) was performed to examine the group relationships, revealing a main effect for Factor 1 [F(1,11)=4.99, $*p<0.05$]. These results indicate that nonprofessional players exhibited higher SCR frequencies than professional players. No significant differences in SCR frequencies were shown between events.

\subsection{Player Response with High Arousal}
\begin{figure}[tb]
      \begin{center}
      \includegraphics[width=\textwidth]{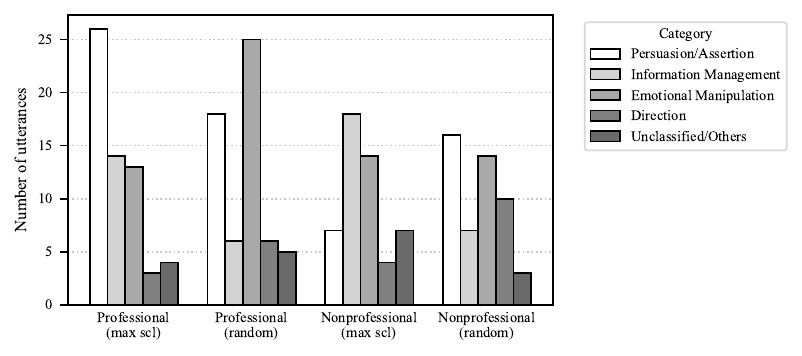}
      \caption{Distribution of communication patterns across professional and nonprofessional players under two conditions: maximum SCL moments and randomly selected first day discussions.}
      \label{fig:results_response}
      \end{center}
\end{figure}

We analyzed the verbal expressions of professional and nonprofessional players under two conditions: during maximum SCL peaks and during randomly selected first-day discussions. The expressions were categorized into five main categories (Table \ref{table:category}). Figure \ref{fig:results_response} shows the distribution of communication patterns under different conditions. The horizontal axis shows combinations of player type (professional / nonprofessional) and conditions (max SCL/random). The vertical axis shows the percentage of patterns in each category. We categorized responses into five main categories: Persuasion/Assertion, Information Management, Emotional Manipulation, Direction, and Unclassified / Other.

As shown in Figure \ref{fig:results_response}, the distribution of communication patterns varies between professional and nonprofessional players, and between high arousal and random discussion periods. The most notable differences appear in the use of persuasion / assessment strategies and information management, particularly during maximum SCL periods.

The expressions of the professional players during the maximum periods of SCL showed a high proportion of persuasion and assertion (43.3\%), followed by information management (23.3\%) and emotional manipulation (21.7\%). During randomly selected discussion periods, the distribution shifted to emotional manipulation (41.7\%) and persuasion/assertion (30\%), with lower frequencies of information management (10\%) and behavioral direction (10\%). When comparing the two conditions, persuasion/assertion increased during high arousal (+13.3\%), while emotional manipulation decreased (-20\%). Information management showed an increase during high arousal states (+13.3\%).

nonprofessional players showed different distributions across conditions. During maximum SCL periods, the expressions were primarily focused on information management (36\%) and emotional manipulation (28\%), with less emphasis on persuasion and assertion (14\%). During random discussion periods, expressions were distributed in persuasion / assessment (32\%), emotional manipulation (28\%), and behavioral direction (20\%). The comparison between conditions shows that information management increased during high arousal (+22\%), while persuasion/assertion decreased (-18\%). Emotional manipulation remained constant (28\%) in both conditions.

The comparison reveals distinct differences in communication patterns between professional and nonprofessional players. During high periods of SCL, professionals predominantly used persuasion / assessment (43.3\%), while nonprofessionals relied more on information management (36\%). The transition patterns from random to max SCL conditions also differed between the groups. Professionals showed an increase in persuasion (+13.3\%), while nonprofessionals demonstrated an increase in information management (+22\%) with a decrease in persuasion / assessment (-18\%). Regarding emotional manipulation, both groups showed different baseline levels during random discussions (professionals: 41.7\%, nonprofessionals: 28\%), with professionals showing greater variation between conditions.

\begin{figure}
    \centering
    \includegraphics[width=\linewidth]{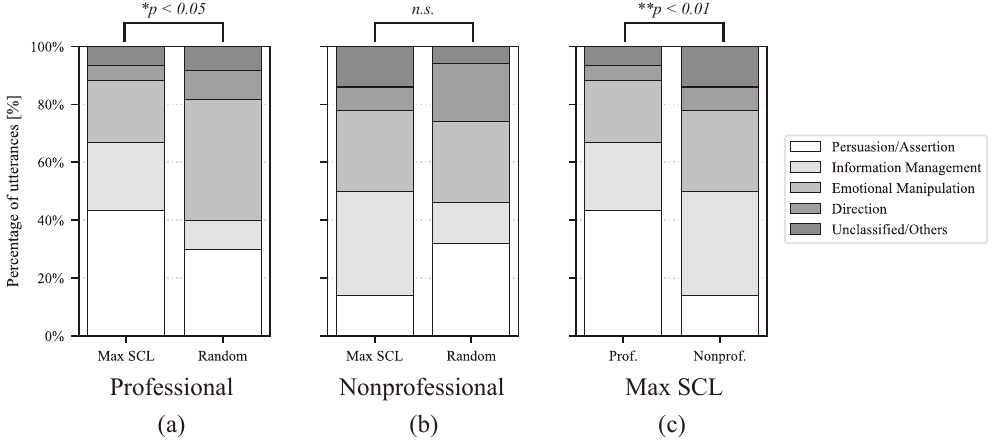}
    \caption{Comarison of communication patterns between conditions and player types. (a) Professional players' patterns in max SCL versus random conditions, (b) Nonprofessional players' patterns in max SCL versus random conditions, (c) Communication patterns between professional and nonprofessional players in max SCL. Statical significance was determined using Mann-Whitney U tests ($^{*}p < 0.05,\ ^{**}p<0.01, n.s.:not\ significant$).}
    \label{fig:comp_patterns}
\end{figure}

Statistical analysis using Mann-Whitney U tests revealed significant differences in the distribution of expression categories. As shown in Figure \ref{fig:comp_patterns} (a), professional players showed significantly different patterns between max SCL and random conditions (U = 1388.5, $p < 0.05$). In contrast, Figure \ref{fig:comp_patterns} (b) shows that nonprofessional players did not show significant differences between max SCL and random conditions (U = 1165.5, $p = 0.56$). Furthermore, Figure \ref{fig:comp_patterns} (c) shows the comparison between professional and nonprofessional players during the maximum SCL periods revealing significantly different communication patterns (U = 1040, $p < 0.01$).

\section{Discussion}

Our analyses revealed the following key findings:
\begin{enumerate}
    \item Physiological responses (electrodermal activity patterns)
    \begin{itemize}
        \item Both player types showed increased arousal during early discussion and voting game phases
        \item Professional players maintained lower SCL values and lower frequencies of SCR
    \end{itemize}
    \item Communication patterns
    \begin{itemize}
        \item Professional players employed persuasion-based approaches during high arousal periods
        \item Nonprofessional players employed information management during high arousal
        \item Stastical analysis showed significant differences in communication patterns between professional and nonprofessional players.
    \end{itemize}
\end{enumerate}
These findings provide insights into how expertise manifests in social deception games in player types. In the following, we discuss how different levels of proficiency in the werewolf game lead to differences in player states and communication strategies.

\subsection{Expertise in Werewolf Games From Persipective on EDA measurement}
The results showed that the normalized SCL were significantly higher during the early stage and the final phase (voting phase) of the game. This pattern was observed in both the professional and nonprofessional groups, indicating a trend independent of the expertise of the player. In nonprofessional games, role claims (declaring one's role, particularly claiming to be the seer) were consistently observed in all trials during the early stage of the Discussion phase. Not only the true seer, but also the werewolf, madman, and villagers made role claims, often leading to situations where all players declared roles. In contrast, professional games exhibited diverse strategic approaches, including selective role claims, casting suspicion on specific players, intentional silence, and information gathering through targeted questioning. In both cases, the strong arousal observed in the early phase reflects its critical importance in setting the game's trajectory. The heightened arousal likely stems from the game's strategic depth, where early-stage actions (role claim patterns in nonprofessional games; tactical information management in professional games) significantly influence the subsequent game development and overall entertainment value.

Significant differences in normalized SCL and SCR frequencies between professional and nonprofessional players showed consistently higher values for nonprofessional players. The SCL results suggest that professional players maintained a state with fewer arousal responses compared to nonprofessionals. Maintaining arousal levels indicates a state without arousal responses, representing a relaxed state or sustained attention \cite{ElectrodermalSystem}. The lower frequency of SCR among professionals, which serves as an indicator of thought transitions \cite{ElectrodermalResponses}, suggests a more stable and focused cognitive processing. This could indicate that professional players maintain more consistent strategic thinking patterns, while nonprofessionals experience more frequent shifts in their tactical considerations and emotional responses to game events.

These differences in emotion dynamics between professional and nonprofessional players highlight different approaches to game strategy and emotional control. Professional players demonstrate the ability to maintain sustained attention while efficiently processing game information, as evidenced by their stable SCL patterns and lower SCR frequencies. In contrast, nonprofessional players show more reactive emotional patterns, with frequent thought transitions suggesting less structured strategic thinking. These findings contribute to our understanding of the expertise in social deception games, where emotional control and strategic consistency appear to be key characteristics of professional play.

\subsection{Communication Pattern Differences Between Professional and Nonprofessional Players}
Analysis of communication patterns revealed distinctive strategies between professional and nonprofessional players. Professional players predominantly used persuasion-based strategies (43.3\% of their expressions), while nonprofessional players relied more on information management, particularly during emotionally charged moments (36\% at maximum SCL). This difference in strategic focus suggests that players at different skill levels approach the game with distinct communication styles, rather than indicating a superior strategy. 

The divergence in strategies becomes particularly apparent during emotionally intensive moments. Although professional players maintained their focus on persuasion even under high arousal, nonprofessional players moved toward information-focused communication. This pattern indicates that players at different skill levels process and respond to emotional pressure differently, adopting different approaches to problem solving in social deception games. These findings suggest that the effectiveness of communication strategies in social deception games may depend on the level of skill of both the speaker and the recipients. The persuasion-based approach of professional players might be particularly effective when interacting with experienced players who can follow complex logical arguments. Conversely, nonprofessional players' focus on information sharing could be more effective in games with less experienced players who prioritize gathering and processing basic game information.

The observed differences in communication patterns have important implications for developing AI players for social deception games. Current AI systems often employ fixed strategies, but our findings suggest the need for adaptive communication approaches that can be adjusted based on the skill level of human players. An effective AI system should be able to switch between persuasion-focused and information-focused strategies, matching its communication style to the characteristics and preferences of its human counterparts. Furthermore, these results highlight the complexity of human experience in social deception games. Rather than following a single optimal strategy, skilled players appear to develop distinctive approaches to the game. This suggests that AI systems aiming to engage effectively in social deception games need to incorporate flexible communication strategies that can adapt to different player skill levels and preferences.

\subsection{Limitation}
Our study has several limitations that should be addressed in future research. First, although we measured electrodermal activity (EDA) using standard equipment, differences in measurement devices and their calibration methods might affect the absolute conductance values. Therefore, our findings should be interpreted in terms of relative changes and patterns rather than absolute conductance values.

Second, this study was conducted in Japan with Japanese participants, which could limit the generalizability of our findings in different cultural contexts. Social deception games such as Werewolf might be played differently in other cultures, and communication strategies may vary based on cultural norms and practices. Another limitation is the binary categorization of participants into professional and nonprofessional players, not including experienced gamers outside theatrical performance contexts. Including a third group with extensive gameplay experience but no professional theatrical training might reveal additional communication strategies. 

Third, our sample size was relatively small, with 60 utterances from professional players (n=7) and 50 from nonprofessional players (n=6). Although this provided sufficient data for our statistical analyzes, the unequal group sizes combined with the small overall sample size may potentially bias results in favor of the one group. A larger and more balanced dataset such as Zhang's work \cite{Werewolf-XL} would be beneficial to validate our findings and potentially uncovering more subtle patterns in communication strategies. However, it should be noted that conducting experiments with simultaneous EDA measurements for five players, especially with professional players, requires significant effort and coordination. Creating a larger more balanced data set would require considerable time and resources.

Finally, while our categorization of communication patterns was executed by a single researcher, which presents a limitation, we took steps to ensure reliability. Future studies could benefit from multiple researchers and a more rigorous validation of the categorization scheme. Our categorization framework relies on existing utterance tags from previous research \cite{AnalysisOfNonVerbalInformation}. In addition, we established categorization criteria (see Appendix A) prior to analysis. These methodological choices support the reliability of our findings.

\section{Conclusion}
This study investigated the differences between professional and nonprofessional players in the Werewolf game through physiological responses and communication patterns. Our key findings revealed distinct characteristics in both physiological responses and communication strategies based on player expertise. First, we found that professional players maintain more stable emotional states during gameplay, as evidenced by their SCL patterns. Although both professional and nonprofessional players showed increased arousal during critical game phases, professionals demonstrated better emotional control and sustained attention. Second, we identified significant differences in communication strategies between professional and nonprofessional players. Professional players consistently used persuasion-based approaches, maintaining this strategy during high emotional arousal. In contrast, nonprofessional players tended to focus on information management, particularly during emotionally intensive moments. These differences suggest that the expertise in social deception games manifests itself not only in emotional control but also in strategic communication choices.

In future work, the patterns identified in this study could be explored for their potential implications in developing more sophisticated AI systems. One promising direction would be investigating how AI systems for social deception games might adapt their communication strategies based on the player's expertise level, potentially switching between persuasion-focused and information-focused approaches. Additionally, the observed relationship between expertise and emotional responses suggests valuable research questions about how humans process and respond to different types of messages. While our current study provides only preliminary insights, these findings open up interesting avenues for research on adaptive communication strategies that consider both logical and emotional aspects of human interaction, which could eventually inform more emotionally intelligent AI systems.

\section*{Acknowledgements}
\noindent
[Omitted for review]
% This work was supported by JSPS KAKENHI Grant Number JP19H04232 and JP25K21365.

% Appendicies
% \input{Include/Appendix_A}

\section*{Declaration of Generative AI and AI-assisted technologies in the writing process}
\noindent
Statement: During the preparation of this work the author(s) used Claude 3.5 Sonnet in order to improve readability and language. After using this service, the author(s) reviewed and edited the content as needed and take(s) full responsibility for the content of the publication.

\bibliographystyle{elsarticle-num-names} 
\bibliography{index}

@article{SuperhumanPokerAI,
author = {Noam Brown  and Tuomas Sandholm },
title = {Superhuman AI for multiplayer poker},
journal = {Science},
volume = {365},
number = {6456},
pages = {885-890},
year = {2019},
doi = {10.1126/science.aay2400},
}

@INPROCEEDINGS{AIWolf1,
  author={Kato, Shoma and Okumura, Tomoya and Toda, Itsuki and Fukui, Takanori and Iwata, Kazunori and Ito, Nobuhiro},
  booktitle={2019 6th International Conference on Computational Science/Intelligence and Applied Informatics (CSII)}, 
  title={Consideration of the Essential Topics for Role Estimation for AIWolf}, 
  year={2019},
  volume={},
  number={},
  pages={72-77},
  doi={10.1109/CSII.2019.00020}}

@article{AIWolfProject1,
  title={Development of Game AI by using Collective Intelligence},
  author={Fujio Toriumi and Michimasa Inaba and Hirotaka Osawa and Daisuke Katagami and Kosuke Shinoda and Hitoshi Matsubara},
  journal={Journal of Digital Games Research},
  volume={9},
  number={1},
  pages={1-11},
  year={2016},
  doi={10.9762/digraj.9.1\_1}
}

@InProceedings{AIWolfProject2,
author="Toriumi, Fujio
and Osawa, Hirotaka
and Inaba, Michimasa
and Katagami, Daisuke
and Shinoda, Kosuke
and Matsubara, Hitoshi",
editor="Cazenave, Tristan
and Winands, Mark H.M.
and Edelkamp, Stefan
and Schiffel, Stephan
and Thielscher, Michael
and Togelius, Julian",
title="AI Wolf Contest --- Development of Game AI Using Collective Intelligence ---",
booktitle="Computer Games",
year="2017",
publisher="Springer International Publishing",
address="Cham",
pages="101--115",
isbn="978-3-319-57969-6"
}

@INPROCEEDINGS{AnalysisOfNonVerbalInformation,
  author={Katagami, Daisuke and Takaku, Shono and Inaba, Michimasa and Osawa, Hirotaka and Shinoda, Kosuke and Nishino, Junji and Toriumi, Fujio},
  booktitle={2014 IEEE International Conference on Fuzzy Systems (FUZZ-IEEE)}, 
  title={Investigation of the effects of nonverbal information on werewolf}, 
  year={2014},
  volume={},
  number={},
  pages={982-987},
  doi={10.1109/FUZZ-IEEE.2014.6891847}}

@INPROCEEDINGS{AnalysisOfNonVerbalInformationForAgentDesign,
  author={Katagami, Daisuke and Kanazawa, Masashi and Toriumi, Fujio and Osawa, Hirotaka and Inaba, Michimasa and Shinoda, Kosuke},
  booktitle={2015 IEEE International Conference on Fuzzy Systems (FUZZ-IEEE)}, 
  title={Movement design of a life-like agent for the werewolf game}, 
  year={2015},
  volume={},
  number={},
  pages={1-7},
  doi={10.1109/FUZZ-IEEE.2015.7338087}}

@INPROCEEDINGS{WerewolfAnalysisWithGazeMotion,
  author={Katagami, Daisuke and Nira, Satoshi},
  booktitle={2017 IEEE Symposium Series on Computational Intelligence (SSCI)}, 
  title={Development of werewolf match system with analysis of human gaze motion}, 
  year={2017},
  volume={},
  number={},
  pages={1-6},
  doi={10.1109/SSCI.2017.8285417}}

@inproceedings{EDAAnalysisInFPSGame,
author = {Drachen, Anders and Nacke, Lennart E. and Yannakakis, Georgios and Pedersen, Anja Lee},
title = {Correlation between heart rate, electrodermal activity and player experience in first-person shooter games},
year = {2010},
isbn = {9781450300971},
publisher = {Association for Computing Machinery},
address = {New York, NY, USA},
doi = {10.1145/1836135.1836143},
pages = {49–54},
numpages = {6},
location = {Los Angeles, California},
series = {Sandbox '10}
}

@inproceedings{AffectiveGaming,
author = {Aggag, Ahmed and Revett, Kenneth},
title = {Affective Gaming: A GSR Based Approach},
year = {2011},
isbn = {9781618040190},
publisher = {World Scientific and Engineering Academy and Society (WSEAS)},
address = {Stevens Point, Wisconsin, USA},
pages = {262–266},
numpages = {5},
location = {Corfu Island, Greece}
}

@article{InterpersonalAutonomicPhysiology,
author = {Richard V. Palumbo and Marisa E. Marraccini and Lisa L. Weyandt and Oliver Wilder-Smith and Heather A. McGee and Siwei Liu and Matthew S. Goodwin},
title ={Interpersonal Autonomic Physiology: A Systematic Review of the Literature},
journal = {Personality and Social Psychology Review},
volume = {21},
number = {2},
pages = {99-141},
year = {2017},
doi = {10.1177/1088868316628405},
note ={PMID: 26921410},
}

@article{DecisionMakingWithEDA,
  title={Anticipatory electrodermal activity and decision making in a computer poker-game.},
  author={Palom{\"a}ki, Jussi and Kosunen, Ilkka and Kuikkaniemi, Kai and Yamabe, Tetsuo and Ravaja, Niklas},
  journal={Journal of neuroscience, psychology, and economics},
  volume={6},
  number={1},
  pages={55},
  year={2013},
  publisher={Educational Publishing Foundation}
}

@article{DeceptiveBehaviorWithEDA,
  title={Behavioral inhibition and electrodermal activity during deception.},
  author={Pennebaker, James W and Chew, Carol H},
  journal={Journal of personality and social psychology},
  volume={49},
  number={5},
  pages={1427},
  year={1985},
  publisher={American Psychological Association}
}

@article{EmotionDynamics,
title = {Emotion dynamics},
journal = {Current Opinion in Psychology},
volume = {17},
pages = {22-26},
year = {2017},
note = {Emotion},
issn = {2352-250X},
doi = {https://doi.org/10.1016/j.copsyc.2017.06.004},
url = {https://www.sciencedirect.com/science/article/pii/S2352250X16302019},
author = {Peter Kuppens and Philippe Verduyn}
}

@article{AlphaGo,
  title={Mastering the game of Go with deep neural networks and tree search},
  author={Silver, David and Huang, Aja and Maddison, Chris J and Guez, Arthur and Sifre, Laurent and Van Den Driessche, George and Schrittwieser, Julian and Antonoglou, Ioannis and Panneershelvam, Veda and Lanctot, Marc and others},
  journal={nature},
  volume={529},
  number={7587},
  pages={484--489},
  year={2016},
  publisher={Nature Publishing Group}
}

@INPROCEEDINGS{ExtendedBDIModel,
  author={Naoyuki, Nide and Takata, Shiro},
  booktitle={2016 IEEE International Conference on Agents (ICA)}, 
  title={Tracing Werewolf Game by Using Extended BDI Model}, 
  year={2016},
  volume={},
  number={},
  pages={7-12},
  doi={10.1109/ICA.2016.014}}

@ARTICLE{Werewolf-XL,
  author={Zhang, Kejun and Wu, Xinda and Xie, Xinhang and Zhang, Xiaoran and Zhang, Hui and Chen, Xiaoyu and Sun, Lingyun},
  journal={IEEE Transactions on Affective Computing}, 
  title={Werewolf-XL: A Database for Identifying Spontaneous Affect in Large Competitive Group Interactions}, 
  year={2023},
  volume={14},
  number={2},
  pages={1201-1214},
  doi={10.1109/TAFFC.2021.3101563}}

@INPROCEEDINGS{HumanLikeWerewolfAgents,
  author={Nakamura, Noritsugu and Inaba, Michimasa and Takahashi, Kenichi and Toriumi, Fujio and Osawa, Hirotaka and Katagami, Daisuke and Shinoda, Kousuke},
  booktitle={2016 IEEE Symposium Series on Computational Intelligence (SSCI)}, 
  title={Constructing a Human-like agent for the Werewolf Game using a psychological model based multiple perspectives}, 
  year={2016},
  volume={},
  number={},
  pages={1-8},
  doi={10.1109/SSCI.2016.7850031}}

@INPROCEEDINGS{DevelopWerewolfAgentswithQLearning,
  author={Wang, Tianhe and Kaneko, Tomoyuki},
  booktitle={2018 Conference on Technologies and Applications of Artificial Intelligence (TAAI)}, 
  title={Application of Deep Reinforcement Learning in Werewolf Game Agents}, 
  year={2018},
  volume={},
  number={},
  pages={28-33},
  doi={10.1109/TAAI.2018.00016}}

@InProceedings{WerewolfModelBasedOnPlayLogAnalysis,
author="Hirata, Yuya
and Inaba, Michimasa
and Takahashi, Kenichi
and Toriumi, Fujio
and Osawa, Hirotaka
and Katagami, Daisuke
and Shinoda, Kousuke",
editor="Plaat, Aske
and Kosters, Walter
and van den Herik, Jaap",
title="Werewolf Game Modeling Using Action Probabilities Based on Play Log Analysis",
booktitle="Computers and Games",
year="2016",
publisher="Springer International Publishing",
address="Cham",
pages="103--114",
isbn="978-3-319-50935-8"
}

@inproceedings{QSensor,
  title={Ambulatory EDA: Comparisons of bilateral forearm and calf locations},
  author={Fedor, Szymon and Picard, Rosalind},
  booktitle={PSYCHOPHYSIOLOGY},
  volume={51},
  pages={S76--S76},
  year={2014},
  organization={WILEY-BLACKWELL 111 RIVER ST, HOBOKEN 07030-5774, NJ USA}
}

@article{ElectrodermalSystem,
  title={The electrodermal system},
  author={Dawson, Michael E and Schell, Anne M and Filion, Diane L and others},
  journal={Handbook of psychophysiology},
  volume={2},
  pages={200--223},
  year={2007}
}

@article{ElectrodermalResponses,
  title={Electrodermal responses: what happens in the brain},
  author={Critchley, Hugo D},
  journal={The Neuroscientist},
  volume={8},
  number={2},
  pages={132--142},
  year={2002},
  publisher={SAGE Publications Sage CA: Los Angeles, CA}
}




% %% For authoryear reference style

% \bibitem[Lamport(1994)]{lamport94}
%   \textit{\LaTeX: a document preparation system},
%   Addison Wesley, Massachusetts,

% \end{thebibliography}
\end{document}